\pdfoutput=1 
\documentclass{JINST}
\let\ifpdf\relax
\usepackage[numbers]{natbib}
\usepackage{amsmath}
\DeclareGraphicsExtensions{.pdf,.png,.jpg}
\newcommand{\vect}[1]{\overrightarrow{#1}}

\title{An introduction to some imperfections of CCD sensors}

\author{P. Astier$^a$\\
\llap{$^a$} LPNHE/IN2P3/CNRS,\\
  UPMC, 4 place Jussieu F75005 Paris, France\\
E-mail: \email{pierre.astier@in2p3.fr}}

\abstract{CCD sensors do not deliver a perfect image of the light they
  receive. Beyond the well known linear image smearing due to
  diffusion of charges during their drift towards the pixel wells,
  non-linear effects are at play in these sensors. We now have ample
  evidence for both a flux-dependent and static image distortions,
  especially but not only, on deep-depleted CCDs. For large surveys
  relying on CCD sensors, these effects should now be taken into
  account when reducing data. We present here a summary of current
  results on sensor characterization and mitigation methods.}


\begin{document}

\section{Introduction}\label{sec:intro}
The first workshop on `` Precision Astronomy with Fully Depleted
CCDs'' was held at Brookhaven National Laboratory in November 2013.
Most of the
contributions\footnote{see \href{http://www.bnl.gov/cosmo2013/}{{http://www.bnl.gov/cosmo2013}}}
discussed sensor effects affecting astronomical measurements using
CCDs. We will here mostly summarize the findings reported then, and
for some aspects, reported since then in the literature. A second iteration of 
the same workshop took place in December 2014 in the same place. New results
where presented there, and we refer interested readers to the program and
proceedings\footnote{see \href{http://www.bnl.gov/paccd2014/}{http://www.bnl.gov/paccd2014}}.

CCD do not map the incoming light along a perfect rectilinear and
periodic grid. The departures of CCD images from this ideal mapping
can be broadly split into two classes: static effects that do not
depend on the charge stored in the device, and dynamic effects that
increase as charge accumulates in the device. A striking example of
static distortions are the so-called tree rings which are visible as
concentric variations of the sensor response to a uniform
illumination. These variations mimic annual growth rings of trees, and
qualify as static because the pattern scales with illumination. We
will discuss shortly the source of these patterns which were observed
only recently on deep-depleted devices.  On the contrary, dynamic
distortions in CCDs vary with illumination, and a typical example is
the variation of the size of Point Spread Function (PSF, i.e. the
response to point sources) with intensity.  Although a sizable PSF
variation was only recently observed (again on deep-depleted CCDs),
there are much earlier reports of phenomena belonging to this class: 
for example a small PSF size variation was predicted more than 40 years ago 
in \citep{Kent73}. About 10 years ago, it was reported that the 
variance in flat-fields does not scale with their average
\cite{Downing06}, at odds with Poisson statistics.
This can also be attributed to dynamic distortions.

The plan of this contribution goes as follows: We first discuss static
distortions, then dynamic ones, and finally provide some outlook.

\section{Static distortions}

By definition static distortions are independent of the illumination
level.  For example a position dependent displacement of astronomical
objects is a static distortion. A historical cause of static
distortions of the images is due to a column or row having a physical
size different from the average. This kind of defect might cause
detectable astrometric shifts (see e.g \cite{Anderson99}, Appendix B
of \cite{Astier13}) and is not specific to some kind of CCD. We do not
know of such defects reported on recent CCDs (in particular the
deep-depleted devices discussed at this workshop), perhaps because
manufacturers have eliminated the causes.  On the contrary, the
defects we discuss in this section seem to be largely specific to
deep-depleted CCDs, or at least they significantly increase with the
device thickness.

\subsection{Tree rings}
The tree ring patterns of flat-field images are observed on most of
the deep-depleted devices. They consist of concentric variations of the
flat-field response to a uniform illumination. The center of the circles
coincides with the center of the ingot from which the sensor was manufactured,
as discussed in \cite{Altmannshofer03}. 
On DECam, these variations
amount to about 0.2\% (y band) to 0.4\% (g band) peak to peak 
\cite{Bernstein_PACCD13}, and the patterns look alike in different bands.
If the flat-field image (with these rings) is used to actually flat-field the science
images, one observes photometry offsets spatially correlated with the
flat-field variations \cite{Bernstein_PACCD13}. This suggests that
these flatfield variations do not originate from conversion efficiency 
inhomogenities but rather from pixel size variations. Indeed, applying
to science images flatfield variations due to pixel size inhomogeneities
{\it degrades } photometric uniformity.   
This hypothesis that the tree rings are due to pixel size variations is strongly
supported by the spatial correlation of astrometry residuals and the
gradient of flat-field variations \cite{Bernstein_PACCD13}. 

The physical picture one can draw is the following: during the growth
of the silicon boule, there are time-dependent variations of silicon purity,
which translate into small static distortions of the
drift field in the sensor\cite{Altmannshofer03,Bernstein_PACCD13}. 
Those distortions in turn displace the
image in a position-dependent way and the divergence of the
displacement field is observed as pixel size variations imprinted on
flat-field images. Since the tree rings have periods above tens
of pixels, one can represent the displacement field as continuous.
Calling $\vect{\delta}$ this displacement field,
(i.e. the average shift in the image plane from the conversion
point to the collection point), and expressing charge conservation, the observed flatfield image $F'$ reads
$$
F'(\vect{x}+\vect{\delta}) = F(\vect{x}) \left|{\mathrm det} \frac{\partial \vect{x}}{\partial (\vect{x}+\vect{\delta})} \right| = F(\vect{x}) \left( 1-div\vect{\delta} + o(\vect{\delta}^2) \right)
$$
where F is the illumination pattern at the conversion point.
The stronger effect at bluer wavelengths is
naturally attributed to their average longer drift path than red
wavelengths (for back-illuminated CCDs), and hence larger transverse shifts. This displacement field ansatz
(due to drift field distortions) also explains the size of the
observed correlations of photometric and astrometric offsets with the
flat-field pattern \cite{Bernstein_PACCD13}. Incidentally, the quality of these correlations
observed on DECam leaves little room (i.e. less that about one
part in a thousand) for flat-field variations due to genuine variations
of sensor efficiency, at least over spatial scales below $\sim$ 50
pixels.

\subsection{Distorted drift fields on sensor edges}
On a CCD device, one might expect distortions of the drift electric field
at a distance from the sensor edges of the order of the sensor thickness.
For thick devices, this can amount to more than ten pixels. Thick deep-depleted
CCDs generically exhibit distortions of the flat-field on the sensor edges,
either a flux deficit (a roll-off, e.g. \cite{OConnor_PACCD13}), or a flux increase 
(``glowing edges'' on DECam, \cite{Bernstein_PACCD13}),
depending on the CCD type. We will generically call roll-off the effect in what 
follows, but what is discussed here applies as well to brighter edge stripes.

One can prove that this response roll-off is due to the electric field
being not perpendicular to the sensor surface, again by relating
astrometric displacements to the flatfield changes. This is done for
the e2v LSST candidates in \cite{OConnor_PACCD13}.  So, the response
roll-off is another manifestation of a displacement field, as tree
rings, and techniques to map tree rings will also capture these
distortions on the sensor edges.

A possible handling of these defects consists of just ignoring the
edges. One can note that tree rings call for a general framework to
handle sensor-induced static astrometric distortions, which could
handle as well the response roll-off on sensor edges, in order to
salvage both astrometric and shape measurements. How much of the
affected edge will be recovered remains unknown today.

The response roll-off on the edges causes spatial charge gradients in
the image. The dynamic distortions we will discuss in next section
tend to smear contrasts, more specifically to cause larger smearing to
larger contrasts.  As a consequence the shape of the roll-off does not
exactly scale with illumination, as was observed on two LSST sensor candidates
\cite{GuyonnetTysonPrivateCommunication}. For
science observations, this means that the handling of the edge
roll-off should in principle depend on the sky background collected in
the image. However, if one implements the correction of dynamic
distortions at the pixel level (as suggested in
\cite{Antilogus14,Gruen15}), the dependence on the sky background of
the response roll-off is accounted for in this correction. The bulk of
the effect still remains to be corrected for, using some mapping of
the displacement field.

\subsection{Static distortions and flat fielding}
When reducing CCD data, it is fairly common to divide every image by
some average of flat-field frames in order to compensate for
efficiency variations. When the spatial variations of flat-field are
dominated by pixel size variations, this is likely to {\it degrade}
the photometric uniformity \cite{Stubbs_PACCD13}. Assuming the contribution
of pixel size variations to the flat-field can be evaluated, one might be
tempted to take those out of the flat-field. This would leave 
spatial variations of the sky background in science images which would
make the sky subtraction inaccurate if not unpractical. So,
over spatial scales smaller than the ones used for sky background evaluation,
the pixel size variations have to be left into the flat-field frame
and their effect on photometry should be corrected for in image catalogs.

\section{Dynamic distortions \label{sec:dynamic}}

Dynamic distortions are characterized by departures from a linear response:
the relation between the image and the illumination (slightly) depends
on the illumination level, even well below the saturation level of the sensor,
and independently of possible non-linearity of the electronic chain.

\subsection{Fat PSF or the brighter-fatter effect} 
The growth of the PSF width with flux (e.g. \cite{Antilogus14}) is a
striking example of dynamic response distortion. The increase in size
reaches $\sim$ 2\% from zero flux to saturation on one of LSST
candidate sensors (the CCD 250 from e2v, 100 $\mu$ thick, 10$\mu$m
pitch). The size increase is smaller ($\sim$ 1\%) on the DECam sensors
(LBNL CCDs, 250 $\mu$m thickness, 13.5 $\mu$m pitch). We do not know of any
failed attempt to observe this effect, even on thinned CCDs
(\cite{Astier13}), where is at only a few per mil level. We do not
know either of any reported evidence of chromaticity of the
effect. All reports are compatible with a slightly steeper increase
(by $\sim$ 20 \%) of the PSF size along the columns
than along the rows. This phenomenon qualifies as non-linear
because the response does not exactly scale with illumination.

A relative increase of 1\% of
the PSF size should be properly addressed for at least two scientific applications
of astronomical imaging : precision PSF photometry of faint vs bright
sources (necessary for example for measuring luminosity distances to
supernovae), and cosmic shear measurements via galaxy ellipticities
(which involve comparing the second moments of faint galaxies and
brighter stars). Qualitatively, the brighter-fatter effect is a smearing
of the image increasing with illumination, and the next section deals
with another behavior that could be described using the same expression.

\subsection{Departure from Poisson fluctuations in uniform exposures}

The non-linearity of the Photon Transfer Curve (PTC)
(e.g. \cite{Downing06}) is some other unexpected behavior of
CCDs. Namely one observes that the spatial variance of uniform
illuminations does not scale with their average, but slightly falls
off, at odds with expectations from Poisson statistics. 
Unsurprisingly, this ``variance deficit'' comes with (mostly) positive
correlations between nearby pixels which decay with spatial
separation.  These correlations also scale with the average
illumination \cite{Downing06,Antilogus14,Lupton_PACCD13}. 

The fact that the variance does not scale exactly with the average
qualifies indeed as non-linearity, as the following sketch shows.  Let
us split a uniform long illumination into smaller time chunks.  If the
device is perfectly linear, the various chunks are independent and
both their averages and variances just add up when assembling the long
exposure. So, a non-linear PTC indicates a response non-linearity,
and one can suspect that 
because of causality, when integrating an image, the late stages of charge
accumulation are influenced by the outcome of earlier stages.

So, the bending of the PTC is non-linear, and corresponds, as the
brighter-fatter effect, to a smearing of the response, increasing with
the exposure level. Since we observe that the statistical
correlations, up to sensor saturation, grow as the average $\mu$ of
the flatfield, we get that the covariance of pixels separated by $i$
rows and $j$ lines vary like $Cov_{ij} \propto \mu V$, where $V$ is
the variance of the flatfield. In a charge-conserving process, this
evolution of covariances comes with a compensating evolution of the
variance (which is hence not exactly Poissonian), because the integral
of the correlation function is conserved by charge redistribution (see
e.g. the appendices of \cite{Downing06,Guyonnet15}). In both
\cite{Downing06,Guyonnet15}, it is also checked that summing the
correlations actually restores the linearity of the PTC.

With $Cov_{ij} \propto \mu V$, i.e.  $Cov_{ij} \propto \mu^2$ to first
order of perturbations, one expects that the variance of uniform
exposures is a quadratic function of their average, as observed
(e.g. \cite{Downing06,Antilogus14,Guyonnet15}). Note that the
quadratic correction to Poisson statistics is not dominant: the
variance deficit is typically of the order of 10 \% at saturation. The
correlations between nearest neighbors are in the percent range at
saturation and rapidly decay to typically $10^{-4}$ at a few pixel
separation.

\subsection{Relating correlations and the brighter-fatter effect}
\begin{figure}[tbp] 
\centering
\includegraphics[width=.7\textwidth]{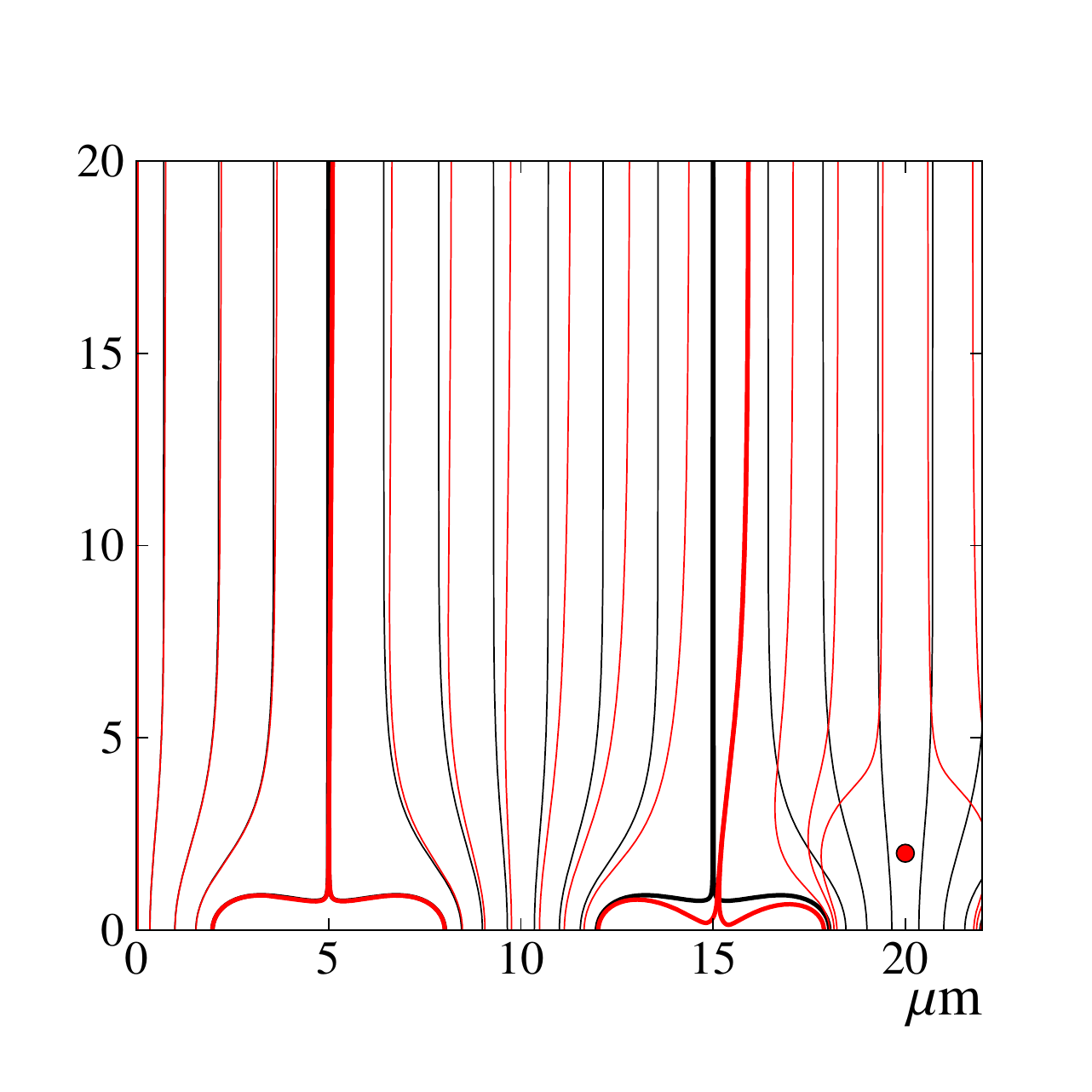}

\caption{Electrostatic computation of field lines in a 100 $\mu$m
  thick, 10 $\mu$m pitch CCD. The clock lines are perpendicular to the
  picture plane, the picture only displays the 20 $\mu$m part that
  contains the clock stripes, and the buried channel has not been
  implemented in the setup .  One set of field lines applies to the
  empty device, and the other applies to the device with 50 000 ke in
  one of the potential wells (red spot). In both instances, the thick
  field lines are the effective pixel boundaries. Note that adding
  this charge actually shrinks the pixel in which it resides, and also
  to a lesser extend its neighbors. Note that the effect of the distortions
  is concentrated in the last microns of the drift and hence essentially 
do not depend
on the conversion point of the photons. The dynamic distortions
are then mostly achromatic, as observed. This figure is borrowed from
  \cite{Guyonnet15}.}
\label{fig:field-lines}
\end{figure}

In \cite{Antilogus14} it is proposed that correlations in uniform
exposures and the brighter-fatter effect share the same origin, namely
the distortions of the electric field sourced by the charges stored in
the potential wells of the sensor, as illustrated by an electrostatic
computation in Fig.  \ref{fig:field-lines}. This assumption is
supported by crude electrostatic calculations, which show that
adjusting the size of field distortions induced by some arbitrary
charge, one can reproduce (to significantly better than a factor of 2)
both the flatfield correlations and how rapidly spot widths increase with flux.
Establishing how stored charges distort the field lines in the
vicinity of the charge buckets requires a precise knowledge of the
geometry and nature of the clock stripes and channel stops.  This is
usually not provided together with sensors, and some of this
information is even regarded as proprietary by sensor vendors.  It is
thus tempting to derive how stored charges affect drift paths from
measured correlations in uniform exposures, in order to infer the
brighter-fatter effect. Note that this simple physical model of
distortions also explains why the observed effects are mostly
achromatic: stored charges affect the electric field mostly in the
vicinity of the pixel wells (see Fig. \ref{fig:field-lines}), and
hence affect all wavelengths equally. We might however expect some
manifestations of chromaticity of correlations at large distances.
The reduction of the drift electric field by charges stored in
the potential wells increases lateral diffusion (see e.g. \cite{Holland_PACCD13}
and references therein), and this
increase is generally a small contribution to the brighter-fatter effect
\cite{Guyonnet15}, but it is chromatic as well. 

If the brighter-fatter effect manifests itself as a
mostly linear increase of the PSF as a function of flux (or peak
flux), one cannot exclude that the effect is more than just
a scaling of size. Identifying the physical mechanism that causes
this apparent increase is the way to safely model the average evolution
of the shape of point-like objects as a function of their flux. Once
this is done, bright stars can safely be used to e.g. unfold the 
finite PSF size from measurements of faint galaxy shapes, as required
by cosmic shear measurements. 

Ref. \cite{Antilogus14} proposes a first order empirical parametrization of
pixel boundary shifts induced by stored charges. This simple model
describes the correlations in uniform exposures linearly rising with
the average content, the quadratic behavior of the PTC\footnote{The
  linear increase of correlations in uniform exposures implies a
  quadratic variation of the PTC, as soon as charge is conserved.},
and the essentially linear rise of the PSF size with peak flux. The
critical test of the model (in fact of any model relating correlations
and the brighter-fatter effect) consists in extracting the model parameters from
flat-fields, and then quantitatively compare the predicted PSF width evolution 
with flux with the measured one.

The model proposed in \cite{Antilogus14} has about twice more
parameters than available correlations to measure. Ref. \cite{Antilogus14},
\cite{Guyonnet15}, and \cite{Gruen15} follow similar strategies in
order to overcome this limitation: they assume that the electrostatic
forces sourced by stored charges decay smoothly with distance. The
details of how smoothness is imposed do not seem to alter
significantly the result. These three works provide evidence that the
brighter-fatter slope can be predicted from correlations in uniform
exposures. Ref. \cite{Guyonnet15} find a 5\% mismatch compatible with the
statistics in use.

\section{Further work and outlook}

We seem to have now physical explanations for both static and dynamic distortions:
the former are due to the drift electric field (in the bulk of the device) 
not being orthogonal to the surface, either because of inhomogeneities of
the bulk properties, or to the electric field being distorted on the edges.
The dynamic distortions are due to the electric field being affected by the 
charges stored in the potential wells. Because the dynamic distortions 
mostly happen near the charge buckets, and the static ones along the whole 
drift path, static distortions are chromatic while the dynamic ones are
mostly achromatic. Note that although the edge roll-off is both
static and dynamic, the separation we are proposing remains pertinent. 

Static distortions are important to map out because they affect both
flux and shape measurements. A detailed study of the static
distortions in DECam is presented in \cite{Bernstein_PACCD13} and
proposes a simple recipe to handle the tree rings: it determines the
displacement field from flat-field inhomogenities, assuming that these
perturbations are entirely due to displacements, and that
displacements follow a radial pattern. This last conditions allows one
to turn a scalar field (the flatfield inhomogenities) into a 2-d
vector displacement field in the image plane. Whether this simple method can
be applied in general is questionable, and a practical method to
separate quantum efficiency variations from a displacement field is
still to be demonstrated. Since the 2-d displacements are due to
transverse drift fields, one can argue that this 2-d displacement
field has a null rotational, which turns it back into a scalar field.
However, one would probably keep this property for a sanity check
rather than reducing the solution space to fields deriving from a
potential.  Incidentally, cosmic shear fields (which are as well
irrotational) are usually not evaluated under this constraint (see
e.g. \cite{Fu08}), but it is rather used as a check.

In order to separate pixel size effects for genuine sensor efficiency
variations, we do not know of a simple method so far. We can think of 
two avenues. In the laboratory, one can illuminate devices with 
spatially varying patterns, and with enough rotations and shifts of 
the same pattern, one can solve for the displacement field, sensor
efficiency, and, if needed, the illumination pattern itself. Several
enterprises along these lines are being developed but the 
proof of concept is still to be delivered. On the sky, one can measure
the displacement field from astrometry (e.g. \cite{Bernstein_PACCD13}),
evaluate the effect on flat-fields and attribute what is left 
in flat fields to sensor efficiency variations. Reasonably, 
projects now assembling their hardware (like LSST) should follow 
both avenues.

Reference \cite{Gruen15} constitutes arguably the first large scale
attempt to infer the brighter-fatter effect from pixel correlations,
and the results are encouraging. One might then regard electrostatic
modelling as a useless complication, especially since this work
detects correlations slopes that change from sensor to sensor, which
are difficult to reduce to variation between sensor batches. In our
opinion, it is still pertinent to develop full realistic electrostatic
models, because we cannot exclude that the sensor parameters 
required for these simulations might be constrained from measurements 
of correlations. Furthermore, since we
expect that some level of chromatism of correlations is eventually
detected, this might further constrain the simulated electrostatic
configurations.

\newpage

\acknowledgments
Both formal and informal discussions at the 2014 and 2013 BNL workshops have been invaluable,
but listing names in this context is almost certainly unfair, and we will not take this risk.
Giving this talk was proposed to us by the workshop organizers, which we warmly thank here,
with a special mention to A. Nomerotski, who acted as the key organizer in both instances
of the workshop.

\def\aap{{\em A\&A}}
\def\apj{ApJ}
\def\apjl{ApJ Lett.}
\def\apjs{ApJ Supp.}
\def\aj{AJ}
\def\prd{Phys. Rev. D}
\def\pasp{PASP}

\bibliographystyle{jhep}
\bibliography{biblio}

\end{document}